\documentclass[preprint,aps,showkeys,superscriptaddress]{revtex4-2}

\usepackage{graphicx}
\usepackage{dcolumn}
\usepackage{bm}
\usepackage{hyperref}
\usepackage{marginnote}
\usepackage{xcolor}
\usepackage{siunitx} 
\usepackage{miller} 

\usepackage[version=4]{mhchem} 

\DeclareSIUnit{\electron}{e^-}

\begin{document}


\title{Direct mapping of electronic orbitals in graphene using electron energy-loss spectroscopy}

\author{M. Bugnet}
\email{mbugnet@superstem.org}
 \affiliation{SuperSTEM Laboratory, SciTech Daresbury Campus, Daresbury WA4 4AD, United Kingdom}
 \affiliation{School of Chemical and Process Engineering, University of Leeds, Leeds LS2 9JT, United Kingdom}
 \affiliation{Univ Lyon, CNRS, INSA Lyon, UCBL, MATEIS, UMR 5510, 69621 Villeurbanne, France}

 \author{M. Ederer}
\affiliation{University  Service  Centre  for  Transmission  Electron  Microscopy, TU Wien, Wiedner Hauptstraße 8-10/E057-02, 1040 Wien, Austria}%

\author{V. K. Lazarov}
\affiliation{Department of Physics, University of York, York YO10 5DD, United Kingdom}%

\author{L. Li}
\affiliation{Department of Physics and Astronomy, University of West Virginia, Morgantown, WV 26506, USA}

\author{Q. M. Ramasse}%
\affiliation{SuperSTEM Laboratory, SciTech Daresbury Campus, Daresbury WA4 4AD, United Kingdom}
\affiliation{School of Chemical and Process Engineering, University of Leeds, Leeds LS2 9JT, United Kingdom}%
\affiliation{School of Physics and Astronomy, University of Leeds, Leeds LS2 9JT, United Kingdom}

\author{S. L\"{o}ffler}
\email{stefan.loeffler@tuwien.ac.at}
\affiliation{University  Service  Centre  for  Transmission  Electron  Microscopy, TU Wien, Wiedner Hauptstraße 8-10/E057-02, 1040 Wien, Austria}%

\author{D. M. Kepaptsoglou}%
\email{dmkepap@superstem.org}
\affiliation{SuperSTEM Laboratory, SciTech Daresbury Campus, Daresbury WA4 4AD, United Kingdom}
\affiliation{Department of Physics, University of York, York YO10 5DD, United Kingdom}

\date{\today}

\begin{abstract}

 The spatial distributions of anti-bonding $\pi^\ast$ and $\sigma^\ast$ states in epitaxial graphene multilayers are mapped using electron energy-loss spectroscopy in a scanning transmission electron microscope. Inelastic channeling simulations validate the interpretation of the spatially-resolved signals in terms of electronic orbitals, and demonstrate the crucial effect of the material thickness on the experimental capability to resolve the distribution of unoccupied states. This work illustrates the current potential of core-level electron energy-loss spectroscopy towards the direct visualization of electronic orbitals in a wide range of materials, of huge interest to better understand chemical bonding among many other properties at interfaces and defects in solids.

\end{abstract}

\keywords{graphene, electronic orbital, core-level spectroscopy, electron energy-loss spectroscopy, fine structures, scanning transmission electron microscopy}
\maketitle


The vast majority of physical and chemical properties of crystalline materials originates from electronic states governing chemical bonding. In addition, defects, interfaces and surfaces have a direct influence on macroscopic material properties. Imaging electronic states, such as chemical bonds at crystal imperfections and discontinuities in real space, is thus of fundamental and technological interest to enable the development of new materials with novel functionalities. While total electronic charge densities have been reconstructed using electron diffraction \cite{zuo1999direct,nakashima2011bonding} or high-resolution imaging \cite{meyer2011experimental} in the transmission electron microscope, and more recently imaged with atomic-scale resolution using four-dimensional scanning transmission electron microscopy (STEM) \cite{muller2014atomic,gao2019real,sanchez2018probing}, the direct observation of individual electronic states has been achieved primarily using scanning tunneling microscopy \cite{repp2006imaging,gross2009chemical,altman2015noncontact}, albeit with surface sensitivity only. 
Electron energy-loss spectroscopy (EELS) in an electron microscope is a spectroscopy technique probing site- and momentum-projected empty states in the conduction band \cite{egerton2011electron}. Following the development of aberration correctors and high stability electron-optics, atomic resolution EELS in the scanning transmission electron microscope (STEM) has become routinely available, leading to elemental (chemical) mapping \cite{bosman2007,kimoto2007element,muller2008atomic}, and providing real-space atomic scale localization of electronic states \cite{bugnet2016real,klie2012observations,nickelate2014atomically,haruta2018atomic,wang2018towards,teurtrie2019atmosphere,gloter2017atomically} using the energy-loss near-edge structure (ELNES) of the spectroscopic signal.

The ELNES, or spectrum fine structure, arising from core-level excitation provides a wealth of information on chemical bonding between atoms, and can be interpreted by first-principles calculations in favorable cases. However, a quantitative interpretation of ELNES maps at atomic resolution requires to also take into account the channeling characteristics of the swift electron beam before and after the inelastic event \cite{tan20112d, lugg2012removing, loffler2013pure, neish2015local, bugnet2016real}, and resulting EELS signal mixing. The appropriate description and/or de-convolution of the electron beam propagation allows for the precise determination of the origin of spatially-resolved variations in fine structures arising from orbital orientation \cite{neishPRB} and localization \cite{loffler2017real}. It has been theoretically predicted that aberration-corrected STEM-EELS should allow for the mapping of electronic orbitals \cite{loffler2013pure}. A first experimental proof of principle was reported through real-space mapping of electronic transitions to \ce{Ti} \textit{d} orbitals in bulk rutile \ce{TiO2}~\cite{loffler2017real}, but thus far, mapping electronic orbitals in real space remains extremely challenging and elusive, be it in bulk crystals or at crystal imperfections and discontinuities.

Graphene, a flagship two-dimensional material with exceptional physical and mechanical properties, has received tremendous scientific interest for potential electronic applications \cite{novoselov2004electric,ak2007rise}. The atomic scale analysis of individual graphene flakes is almost exclusively achieved in top surface view, thus enabling a path to probe single atom chemical bonding \cite{suenaga2010atom,zhou2012direct,ramasse2013probing} and phononic response \cite{hage2020single}. The chemical bonding in graphene can be described as in-C-plane ($\sigma$) and orthogonal out-of-C-plane ($\pi$) covalent bonds. The ELNES of the C-K edge therefore represents the excitation of core states probing in-C-plane $\sigma^\ast$ (1\textit{s} $\rightarrow$ 2\textit{p$_{x,y}$}) orbitals and out-of-C-plane $\pi^\ast$ (1\textit{s} $\rightarrow$ 2\textit{p$_{z}$}) orbitals, as illustrated schematically in Fig. S1. While $\pi^\ast$ state distributions around nitrogen and boron dopants in monolayer graphene have been evidenced from the ELNES \cite{kepaptsoglou2015electronic}, the prospect of mapping orbitals at vacancies and nitrogen dopants in a single graphene sheet has been explored theoretically only a few years ago \cite{pardini2016mapping}. Nevertheless, even if this is intuitively the appropriate direction to observe individual in-C-plane $\sigma$ bonds, inelastic channeling computations show that the STEM-EELS mapping of $\sigma^\ast$ orbitals in top surface view in pristine graphene is not possible due to symmetry constraints \cite{pardini2016mapping,PhysRevLett.126.094802}. The observation of graphene layers in side view, however, provides a pathway to directly visualizing the distribution of $\pi^\ast$ states at the atomic scale using STEM-EELS. While the description of the atomic scale distribution of out-of-C-plane $\pi^\ast$ and in-C-plane $\sigma^\ast$ states may appear simple enough from a chemical bonding perspective, experimental evidence using STEM-EELS is lacking. Moreover, considering the aforementioned subtle effects associated with the localization of the EELS signal, due for instance to channeling of the incident electron beam, the interpretation of energy-filtered real-space maps can be very complex and must be validated through careful numerical work. 

In this letter, real-space $\pi^\ast$ and $\sigma^\ast$ orbital mapping in epitaxial graphene multilayers is achieved in side view, combining state-of-the-art high spatial and energy resolution STEM-EELS with inelastic channeling calculations. The interpretation of the spatial distribution of orbital signals, based on the excellent agreement between computed and experimental data, highlights the successful direct mapping of the $\pi^\ast$ state distribution at atomic resolution in the transmission electron microscope. The theoretical approach provides a powerful platform to determine the origin of the energy-filtered signal.

\begin{figure}
    \centering
    \includegraphics[width=8.6cm]{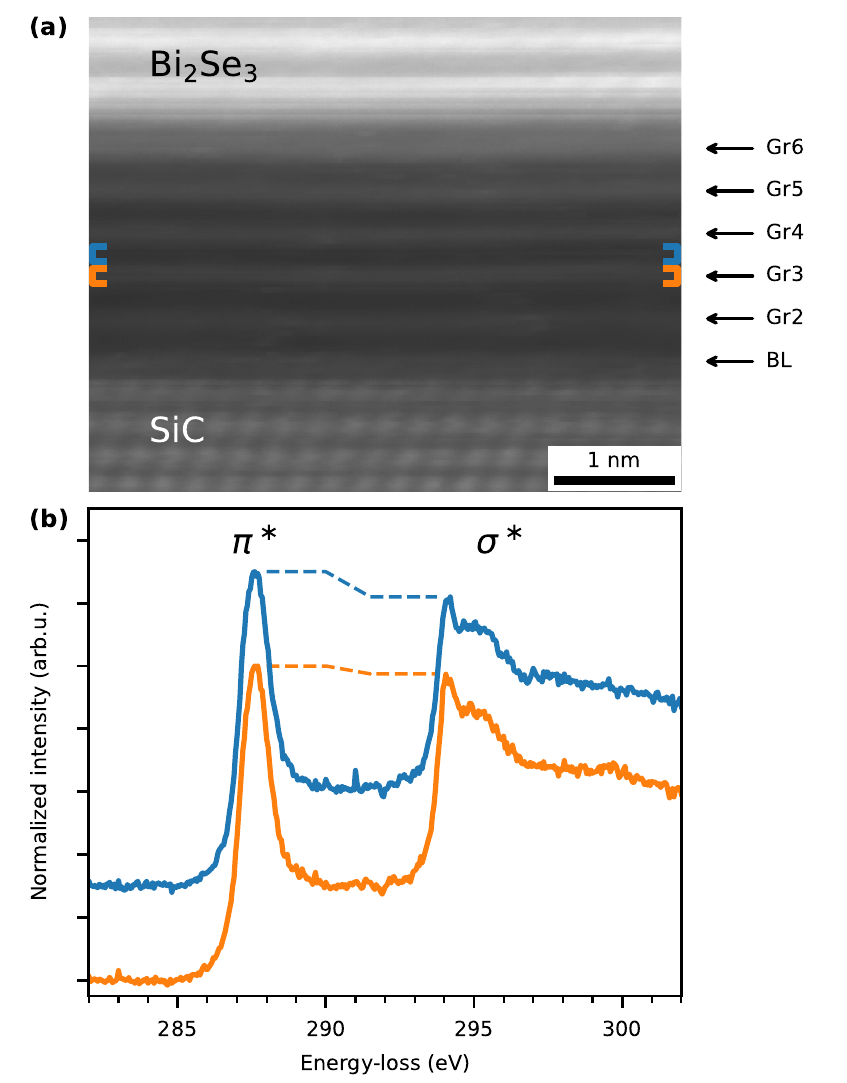}
    \caption{(a) High resolution STEM-HAADF image of a six layer epitaxial graphene assembly, grown on \ce{6H-SiC} and topped with \ce{Bi2Se3}, simultaneously acquired with core-loss EELS data. (b) \ce{C}-K edge spectra corresponding to in-\ce{C}-plane (solid orange line) and between layers (solid blue line), as indicated in (a). The spectra are integrated over the width of the whole image, presented after background subtraction, and shifted vertically for visualization.
}
    \label{fig1}
\end{figure}


\begin{figure*}
    \centering
    \includegraphics{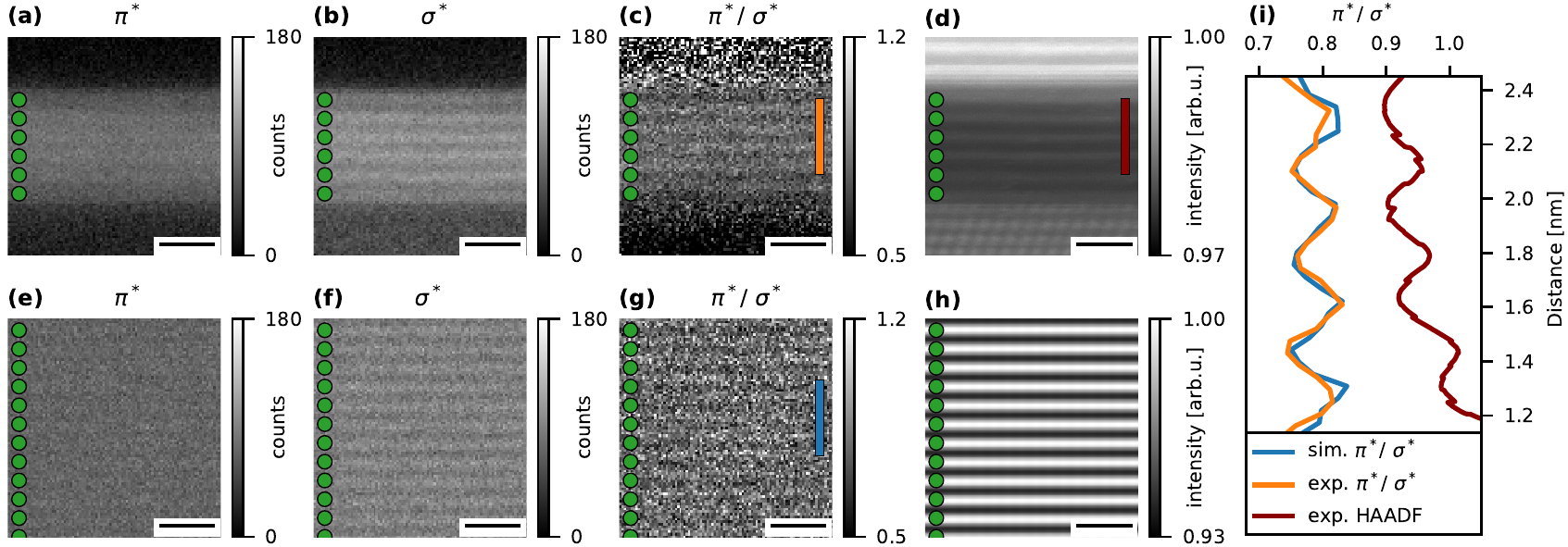}
    \caption{(a, b, c, d) Experimental $\pi^\ast$, $\sigma^\ast$,  $\pi^\ast$/$\sigma^\ast$ maps, and HAADF image, respectively. (e, f, g, h) Theoretical $\pi^\ast$, $\sigma^\ast$,  $\pi^\ast$/$\sigma^\ast$ maps with shot noise, and ADF image, respectively. The position of atomic planes from the HAADF is indicated with green circles. (i) $\pi^\ast$/$\sigma^\ast$ profiles from (c, g), and HAADF intensity integrated in the range indicated by the vertical blue and orange bars in (c) and (g). All scale bars indicate \SI{1}{\nano\meter}.}
    \label{fig2}
\end{figure*}

The epitaxial graphene/\ce{SiC} specimen was synthesized by thermal decomposition of \ce{SiC} at \SI{1300}{\celsius} in ultrahigh vacuum (UHV). For completeness, and as shown in Fig. \ref{fig1}a, we note that a thin capping film of \ce{Bi2Se3} was additionally deposited on top of the graphene layers by molecular beam epitaxy at \SIrange[range-phrase = --, range-units = single]{275}{325}{\celsius} \cite{Liu2014}. This specimen was selected due to the convenient cross-section geometry of the graphene layers; the properties and electronic structure of interfaces with the \ce{6H-SiC} substrate and the \ce{Bi2Se3} capping film are the subject of separate studies and not discussed here. This results in a structure comprising of a so-called graphene 'buffer layer' (BL) in contact with the underlying SiC substrate, capped with a number of layers of 'epitaxial' graphene (here five such layers are seen on Fig.\,\ref{fig1}a), whose macroscopic properties are known to be nigh-on identical to those of free-standing graphene \cite{berger2006electronic,nicotra2013delaminated}.  

The cross-section STEM lamellae were prepared by focused ion beam milling. The thickness of the specimen in the regions of investigation was evaluated to $\sim\SI{25}{\nano\meter}$ by Fourier-Log deconvolution of low-loss EELS spectra ~\cite{egerton2011electron}. The STEM-EELS experiments were carried out using a Nion HERMES microscope, equipped with a high-energy-resolution monochromator, a C\textsubscript{s} aberration corrector up to the 5\textsuperscript{th} order, a Gatan Enfinium spectrometer, and operated at \SI{60}{\kilo\volt}. The convergence and collection semi-angles were \SI{30}{\milli\radian} and \SI{66}{\milli\radian}, respectively. The specimen was oriented in the \hkl[1 0 -1 0] zone axis of \ce{SiC}, corresponding to the \hkl[2 1 -3 0] zone axis of graphene (see Fig. S1). The C-K edge was acquired with a \SI{1.1}{\angstrom} probe size and a step of $\sim\SI{0.3}{\angstrom}$, providing high spatial sampling while preserving the specimen from electron beam damage. The monochromator slit width was adjusted to provide an effective energy resolution of $\sim\SI{100}{meV}$, as measured at the zero-loss peak full width at half maximum. While not the highest achievable resolution on the instrument, these conditions provided a good compromise of beam current (given the chosen probe size) while still being narrower than expected spectral features.
The presented STEM-EELS dataset was acquired from a region of 4.8 $\mathrm{\times}$ \SI{3.9}{\nano\meter}$^{2}$, with a sampling of 110 $\mathrm{\times}$ 88 pixels$^{2}$. Subpixel-scanning (16 $\mathrm{\times}$ 16) was employed, hence leading to a 1760 $\mathrm{\times}$ 1408 pixels$^{2}$ simultaneously acquired HAADF image in Fig. \ref{fig1}a. The experimental EELS maps and HAADF image in Fig. \ref{fig2} are directly cropped from a 88 $\mathrm{\times}$ 88 pixels$^{2}$ region in the original dataset. The dwell time was 0.2s, at a dispersion of 0.05 eV/pixel. The experimental maps were obtained after background extraction (modelled with a power-law function), and energy filtering with a \SI{2}{\electronvolt} window for $\pi^\ast$ and $\sigma^\ast$ states.

The spatial variations of the \ce{C}-K ELNES are highlighted in Fig. \ref{fig1}b, where spectra corresponding to in-C-plane (solid orange line) and out-of-C-plane (solid blue line) are displayed. It is noteworthy that the instrumental broadening of the electron source is narrower than the intrinsic linewidth of the fine structures. This is expected to facilitate orbital mapping since the spectral features are not limited by the energy resolution of the electron source but by physical phenomena linked to, \textit{e.g.}, the excited state lifetime broadening, core-hole screening, or other multi-electronic interactions. The $\pi^\ast$ and $\sigma^\ast$ fine structures are in good agreement with existing work on free standing graphene layers~\cite{nicotra2013delaminated,palacio2015atomic}, with a sharp excitonic feature visible around \SI{294.5}{\electronvolt}. Although the edges overall look comparable in- and out-of-C-plane, the $\pi^\ast$ intensity increases noticeably between the epitaxial graphene layers (out-of-C-plane). This behavior was systematically observed and is characteristic of all \ce{C}-K near-edge structures between the epitaxial graphene layers Gr2--Gr6 (see Fig. S2). The ELNES of the BL and between the graphene BL and Gr2 are influenced by significant covalent bonding between the graphene BL and \ce{SiC}~\cite{nicotra2013delaminated}, and thus are not considered here.
In a first approximation, the spectral variations observed for the Gr2--Gr6 graphene layers can be related to the simple picture of out-of-C-plane delocalization of $\pi^\ast$ states, in contrast to the in-C-plane nature of $\sigma^\ast$ states. Indeed, the lobes of antibonding $\pi^\ast$ orbitals are lying in between \ce{C} planes, whereas the $\sigma^\ast$ orbitals are contained essentially within the graphene planes, as shown schematically in Fig. S1. Nevertheless, the magnitude of this out-of-C-plane delocalization measured by fine structure mapping, and the ability to spatially distinguish $\pi^\ast$ from $\sigma^\ast$ states using a convergent electron-probe in STEM-EELS are non-trivial.

\begin{figure*}
    \centering
    \includegraphics{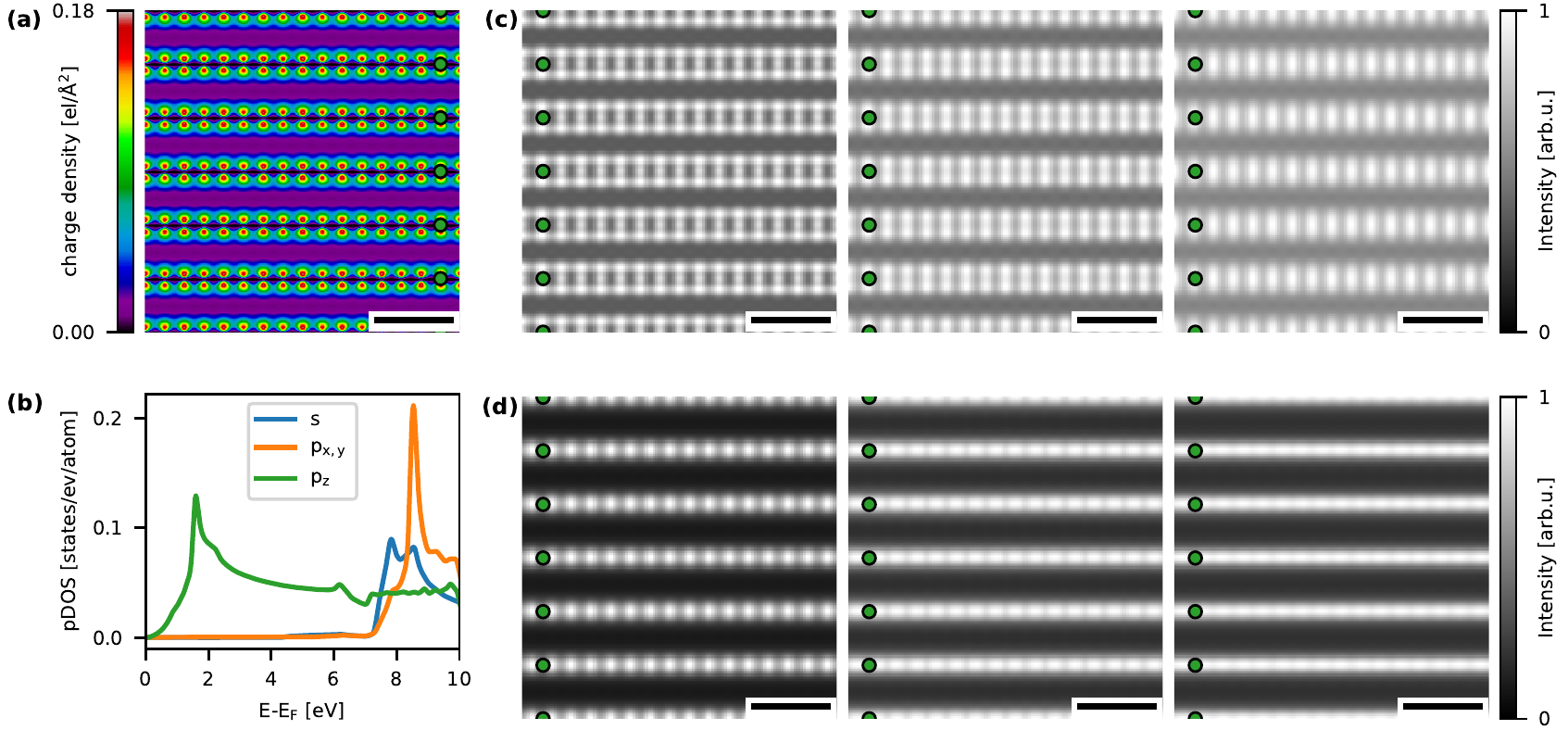}
    \caption{(a) Charge density. (b) Projected density of states in graphite. The z-axis corresponds to the crystallographic c-axis of graphite, perpendicular to the carbon layers. (c)~$\pi^\ast$ map for projected thicknesses of \SI{0.43}{\nano\meter} (left), \SI{12.8}{\nano\meter} (middle), and \SI{25.6}{\nano\meter} (right). The position of atomic planes is indicated with green circles.  (d) $\sigma^\ast$ maps for the same thicknesses. All maps are shown without noise and instrumental broadening. All scale bars indicate \SI{5}{\angstrom}.
}
    \label{fig3}
\end{figure*}

In order to rationalise experimental findings, we carried out extensive numerical calculations of the fine structure maps. The effect of the graphene--\ce{SiC} interface, partially influenced by covalent bonding, and of the graphene--\ce{Bi2Se3} interface on orbital mapping are beyond the scope of this work, therefore a structure made exclusively of graphene layers was considered for inelastic channeling calculations. For simulating the elastic electron propagation both before and after the inelastic scattering events, the multislice algorithm \cite{kirkland1998, cowley1957} was used. For the inelastic interaction between the probe beam and the sample electrons, we calculate the mixed dynamic form factor \cite{loffler2013pure,loffler2013phd}, based on density functional theory data obtained with WIEN2k \cite{blaha2020}.  All simulated STEM-EELS maps were calculated with the same parameters (acceleration voltage, convergence/collection angle, orientation, sampling, etc.) as used in the experiments.
For Fig.~\ref{fig2}, the simulated ideal maps were blurred using a Gaussian filter with a standard deviation of \SI{1.1}{\angstrom} to mimic instrumental broadening due to partial coherence of the electron source \cite{U_v115_i0_p21}. Subsequently, shot noise was added based on the experimental noise characteristics, which were evaluated from the electron intensity in the experimental maps; $\pi^\ast$:~\SI[per-mode=symbol]{31676.4}{\electron\per\nano\meter^2}, $\sigma^\ast$:~\SI[per-mode=symbol]{40026.5}{\electron\per\nano\meter^2}.

The experimental $\pi^\ast$ and $\sigma^\ast$ maps, shown in Figures \ref{fig2}a and \ref{fig2}b, respectively, both display higher intensity where the \ce{C} planes are located. The localization of the $\sigma^\ast$ states on the \ce{C} planes is expected. For the $\pi^\ast$ states (which one might expect to be stronger between the \ce{C} planes), the apparent, counter-intuitive localization on the planes can be explained by channeling effects of the electron beam. These observations are confirmed in the computed maps obtained by inelastic channeling simulations in Figures \ref{fig2}e and \ref{fig2}f.

Rather than analyzing the absolute intensities, we investigate the ratio between the $\pi^\ast$ and the $\sigma^\ast$ intensities as shown in Figures \ref{fig2}c and \ref{fig2}g, as a way to normalize the $\pi^\ast$ intensity variations.
The ratio is maximized between the \ce{C} planes in these maps, as exemplified by the vertical $\pi^\ast$/$\sigma^\ast$ profiles plotted \textit{versus} the HAADF intensity in Fig. \ref{fig2}i. HAADF intensity minima coincide with $\pi^\ast$/$\sigma^\ast$ intensity profile maxima, which are almost exactly equidistant from two graphene layers. The visual agreement between the calculated and experimental $\pi^\ast$, $\sigma^\ast$ and $\pi^\ast$/$\sigma^\ast$ maps is supported by the remarkable overlap of the calculated and experimental $\pi^\ast$/$\sigma^\ast$ line profiles. This successful reproduction of the experimental data underlines the robustness of the inelastic channeling calculations performed in this work to interpret the experimental spectral data. Most importantly, this result provides an undeniable proof that the contrast obtained from $\pi^\ast$ and $\sigma^\ast$ real-space fine-structure maps at high resolution does match the expected localization of corresponding unoccupied electronic orbitals. It also highlights that beyond the atomic site where core-level excitation takes place, the localization of the $\pi^\ast$ and $\sigma^\ast$ orbitals in two-dimensional maps is intimately linked to the channeling of the electron beam, and is thus strongly affected by the specimen projected thickness \cite{hovden2012channeling}.

While the channeling of the swift electron beam primarily depends on the alignment of the electron beam path with the atomic columns, the projected thickness also strongly modifies the atomic-scale contrast in fine structure maps. To evaluate the influence of the projected thickness on the expected $\pi^\ast$ and $\sigma^\ast$ orbital contrast, we performed inelastic channeling calculations considering specimens with different thicknesses: \SI{0.43}{\nano\meter} (a single graphene unit cell), \SI{12.8}{\nano\meter}, and \SI{25.6}{\nano\meter}. The latter corresponds to the estimated thickness of the TEM lamella considered experimentally. The $\pi^\ast$ and $\sigma^\ast$ maps corresponding to these projected thicknesses under ideal conditions (no noise, no instrumental broadening, etc.) are presented in Fig. \ref{fig3}c and \ref{fig3}d, respectively.
The $\pi^\ast$ map of the thinnest specimen displays lobes outside the \ce{C} planes, in agreement with the $\pi^\ast$ charge density in Fig. \ref{fig3}a. Additional intensity is also visible on the \ce{C} columns, and becomes more prominent for larger and more realistic projected thicknesses. At a thickness of \SI{25.6}{\nano\meter}, the intensity of the $\pi^\ast$ states on the \ce{C} columns is stronger than outside the \ce{C} planes, in agreement with the experimental and computed $\pi^\ast$ maps in Fig. \ref{fig2}. For all thicknesses, it is noteworthy that the intensity in the $\pi^\ast$ maps is expected to fade out beyond $\sim\SI{1}{\angstrom}$~away from the \ce{C} planes. The intensity in the $\sigma^\ast$ maps is, as expected, exclusively contained within the \ce{C} planes, and peaked on the \ce{C} columns. In addition, it is noteworthy that the atomic resolution contrast is smoothed out with increasing thickness. It should be noted that the $\sigma^\ast$ maps also contain some intensity from \textit{p$_{z}$} states, \textit{i.e.}, states with $\pi^\ast$ symmetry, as shown in the PDOS in Fig. \ref{fig3}b.

These simulated fine structure maps, in which the elastic channeling conditions of the electron beam were taken into account, highlight the fact that the specimen thickness must be considered carefully to interpret STEM-EELS orbital mapping experiments successfully. Halving the projected thickness down to \SI{12.8}{\nano\meter} is expected to lead to a result similar to the current experimental thickness of \SI{25.6}{\nano\meter}. The direct comparison of $\pi^\ast$  orbital maps with the $\pi^\ast$ charge density plot in Fig. \ref{fig3}a is not reasonable for the experimental thickness considered, nor even for 12.8 nm, but only for an unrealistically small thickness of the order of 0.43 nm. Therefore, it is suggested that smaller projected thickness will only be meaningful below few nm to provide better visualization of electronic orbitals using STEM-EELS in the present case. The noise level is also a major hurdle to overcome, and is clearly visible when comparing the $\pi^\ast$ and $\sigma^\ast$ maps in Figures \ref{fig2}e (shot noise added) and \ref{fig3}c (no shot noise), and Figures \ref{fig2}f (shot noise added) and \ref{fig3}d (no shot noise), respectively. It is expected that orbital mapping in STEM-EELS might benefit from a new generation of detectors with improved sensitivity and lower noise level \cite{PLOTKINSWING2020113067,cheng2020performance}.

In conclusion, the spatial distribution of anti-bonding $\pi^\ast$ and $\sigma^\ast$ orbitals in epitaxial graphene multilayers was mapped successfully by electron energy-loss spectroscopy in the aberration-corrected scanning transmission electron microscope. Inelastic channeling calculations unambiguously reproduce the experimental $\pi^\ast$ and $\sigma^\ast$ orbital maps with high level of accuracy, and demonstrate the decisive effect of the specimen thickness on the orbital mapping capabilities in graphene. The real space visualization, at atomic resolution, of unoccupied electronic states with different symmetry defines a pathway to better understand chemical bonding at interfaces and defects in solids. This is particularly relevant to foster defect engineering and tune the  properties of solids for a wide range of promising applications where physical and chemical phenomena occur at surfaces (\textit{e.g.}, photocatalysis) or interfaces (\textit{e.g.}, spintronics). This work further illustrates the potentiality of orbital mapping using STEM-EELS especially considering recent major improvements in detector performance.

\begin{acknowledgments}
The electron microscopy work was supported by the EPSRC (UK). SuperSTEM Laboratory is the EPSRC National Research Facility for Advanced Electron Microscopy. The authors would like to thank Hitachi High Technologies Corporation (UK \& Japan) and Orsay Physics S.A. \& Tescan Inc. for the preparation of FIB lamellae. M.B. is grateful to the SuperSTEM Laboratory for microscope access, and to the School of Chemical and Process Engineering at the University of Leeds for a visiting associate professorship and financial support. M.E. and S.L. acknowledge funding from the Austrian Science Fund (FWF) under grant nr. I4309-N36.  L.L. acknowledges funding from US National Science Foundation under grant No. EFMA-1741673.

\end{acknowledgments}

M.B. and M.E. contributed equally to this work.

\bibliographystyle{apsrev4-1}

%

\pagebreak
\section{Supplementary Information to Direct mapping of electronic orbitals in graphene using electron energy-loss spectroscopy}

\renewcommand{\thefigure}{S1}
\begin{figure*}[h]
    \centering
    \includegraphics[width=16cm]{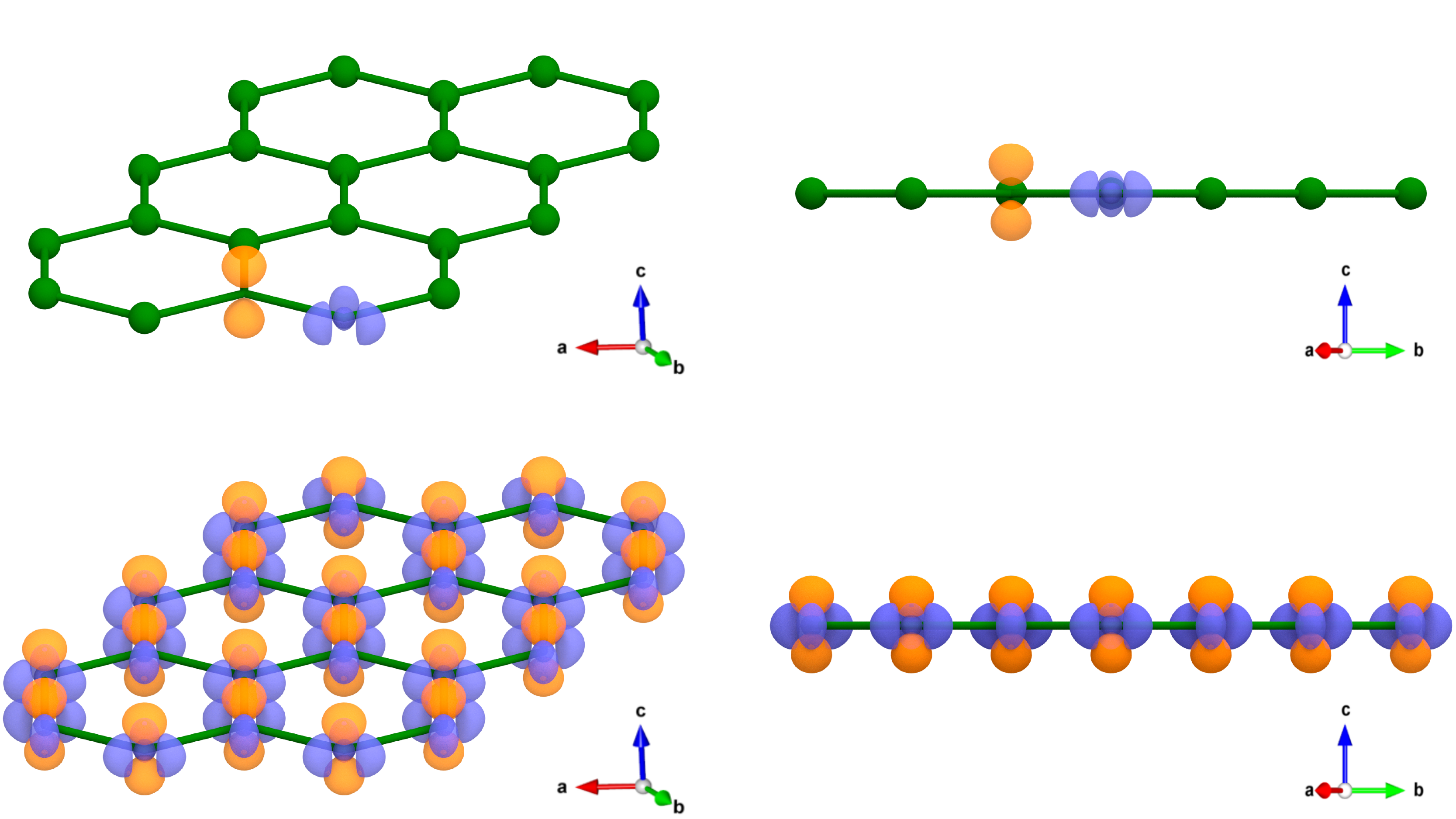}
    \caption{Structural model of graphite used for inelastic channelling calculations, and superimposed schematic representation of $\pi^\ast$ (orange) and $\sigma^\ast$ (blue) electron densities obtained from density functional theory. The inelastic channeling calculations were performed with the graphite layers oriented in the \hkl[2 1 -3 0] zone axis, as shown on the right.
}
    \label{figS1}
\end{figure*}

\renewcommand{\thefigure}{S2}
\begin{figure*}
    \centering
    \includegraphics[width=16cm]{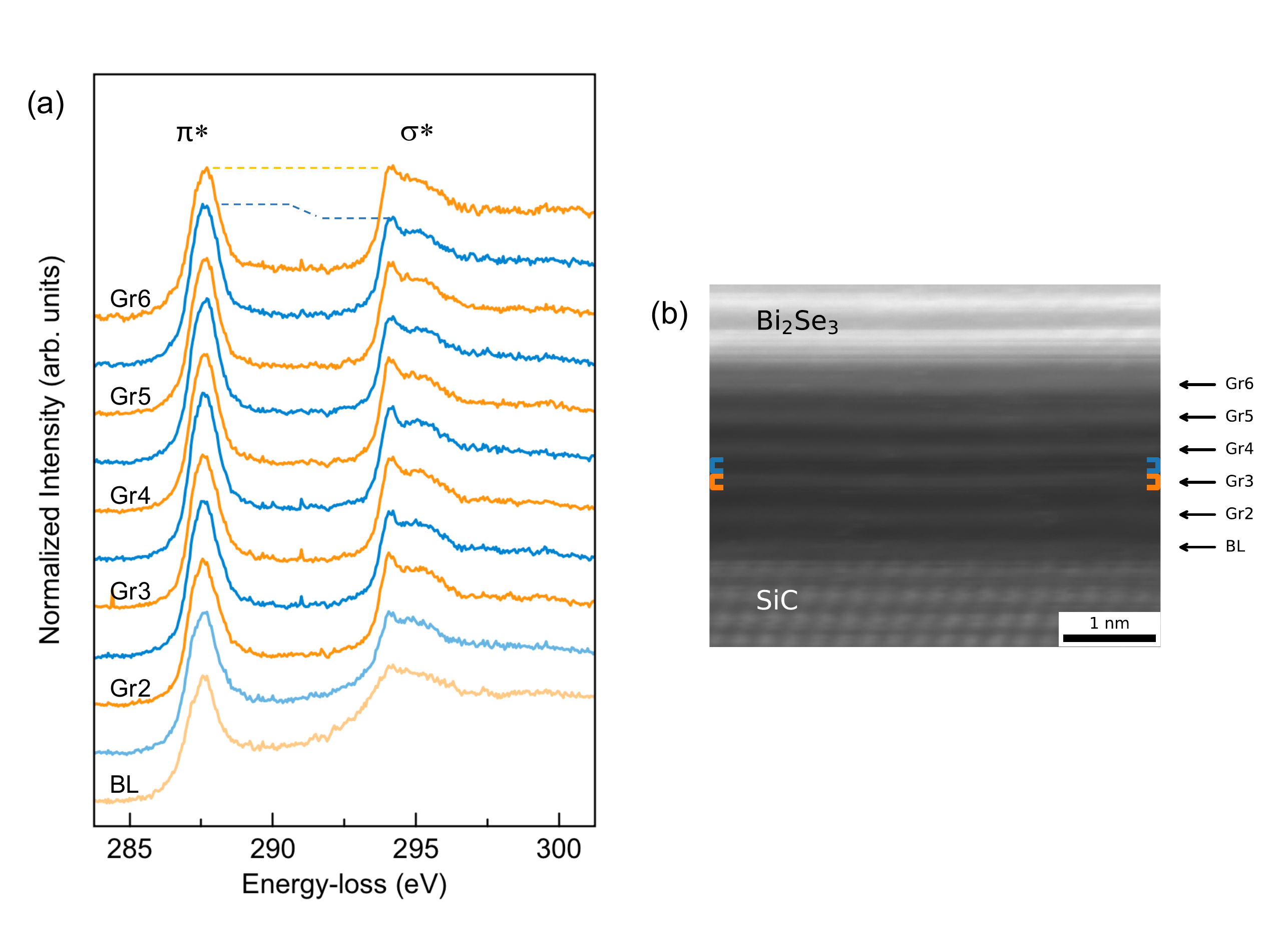}
    \caption{(a) C-K edge spectra corresponding to in-C-plane (orange) and out-of-C-planes (blue) of a six layer epitaxial graphene assembly, grown on \ce{6H-SiC} and topped with \ce{Bi2Se3}, as indicated in the simultaneously acquired HAADF image in (b). The spectra are integrated over the width of the whole image, presented after background subtraction, and shifted vertically for visualization. Relative to $\sigma^\ast$, the $\pi^\ast$ intensity is higher in between the graphene layers. The ELNES of the BL is influenced by covalent bonding between the BL and the Si-terminated SiC; consequently both spectra corresponding to the BL and in between BL and Gr2 do not follow the $\pi^\ast$/$\sigma^\ast$ trend.
}
    \label{figS2}
\end{figure*}

\end{document}